\documentclass[aps,twocolumn,superscriptaddress,showpacs]{revtex4}
\usepackage[dvips]{graphicx}

\renewcommand{\vec}[1]{\mathbf{#1}}
\newcommand{\eqref}[1]{(\ref{#1})}
\renewcommand{\(}{\left(}
\renewcommand{\)}{\right)}

\begin{document}
  
  \title{Strong magnetohydrodynamic turbulence with cross helicity}
  \author{Jean Carlos Perez}
  \author{Stanislav Boldyrev}
  \affiliation{Department of Physics, University of Wisconsin at Madison, 1150
    University Ave, Madison, WI 53706, USA}
  \date{\today}

  \begin{abstract}
    
    Magnetohydrodynamics (MHD) provides the simplest description of
    magnetic plasma turbulence in a variety of astrophysical and laboratory systems.  
    MHD turbulence with nonzero cross helicity is often called
    \emph{imbalanced}, as it implies that the energies of Alfv\'en
    fluctuations propagating parallel and anti-parallel the background
    field are not equal. Recent analytical and numerical studies have
    revealed that at every scale, MHD turbulence consists of regions of
    positive and negative cross helicity, indicating that such turbulence
    is inherently locally imbalanced. In this paper, results from
    high resolution numerical simulations of steady-state incompressible
    MHD turbulence, with and without cross helicity are presented.  It is argued that the inertial range 
    scaling of the energy spectra ($E^\pm$) of fluctuations moving in 
    opposite directions is independent of the amount of cross-helicity. 
    When cross helicity is nonzero, $E^+$ and $E^-$ maintain the same scaling, 
    but have differing amplitudes depending on the amount of cross-helicity.
    
  \end{abstract}

  \maketitle

  \section{Introduction}
  Magnetohydrodynamic (MHD) equations govern the dynamics of a 
  plasma when the spatial scales of interest greatly exceed
  any intrinsic plasma scale. The MHD model is relevant in understanding
  waves, instabilities, and turbulence in a variety of 
  astrophysical systems. For instance, magnetohydrodynamic (MHD)
  turbulence plays a key role in the theoretical modeling of the spectra
  of velocity and magnetic fluctuations in the Solar Wind
  \citep[e.g.,][]{coleman66,coleman68,belcher71,marsch91,goldstein95} as
  well as electron density fluctuations in the interstellar medium
  \citep[e.g.,][]{armstrong95,lithwick01}.

  Theoretical modeling of the spectra in strong MHD turbulence has
  developed on different fronts: phenomenology, statistical closures and
  more recently, high-resolution numerical simulations. The first
  phenomenological model, a la Kolmogorov~\cite{K41,obukhov41}, of the energy spectrum was
  developed independently by Iroshnikov~\citep{iroshnikov63} and
  Kraichnan~\citep{kraichnan65}. In these works they realized that the
  magnetic field of the large scale eddies acts in much the same way as
  a guide field that supports smaller scale Alfv\'enic fluctuations, and
  the turbulent energy transfer takes place as the result of many
  cumulative collisions between counter-propagating Alfv\'en wave packets
  traveling along the local magnetic field.  One deficiency of this
  phenomenology is that it is based on the assumption of an isotropic
  spectral transfer, in contradiction with more recent results that reveal
  the anisotropic character of MHD
  turbulence~\cite[e.g.,][]{montgomery81,shebalin83}. To account for anisotropy in the
  strong turbulence regime, Goldreich and Sridhar~\citep{GS95},
  hereafter referred as GS, proposed a new phenomenology in which they
  introduced the so-called {\em critical balance} condition, stating
  that the turbulence is strong when the time scale for wave propagation
  along the magnetic field matches the nonlinear interaction
  time. In the past decade, constantly increasing massively parallel computer
  clusters have become large enough to allow one to simulate the
  inertial range spectra of strong MHD turbulence, leading to an
  extensive volume of work addressing universal scaling laws in MHD
  turbulence \citep[e.g.,][]{cho00,maron01,haugen03,biskamp,muller05,
  mason06,mason08,mininni07,beresnyak08,perez08,perez09}.  These
  simulations have motivated new phenomenological models and have
  resulted in a renewed interest in the fundamentals of MHD
  turbulence. 

  One aspect of MHD turbulence, which has recently drawn significant
  attention is the role that cross-helicity plays in the turbulent
  cascade~\citep{lithwick07,chandran08,beresnyak09,perez09,podesta10,perez10}. Both total energy $E=\frac{1}{2}\int (v^2 +b^2)d^3 x$ and total cross-helicity $H^c=\int ({\bf v}\cdot {\bf b})d^3 x$ are conserved by nonlinear interactions; where ${\bf v}$ and ${\bf b}$ are velocity and magnetic fields. In MHD turbulence both energy and cross helicity cascade from large to
  small scales as a result of the nonlinear interaction of
  counter-propagating Alfv\'en fluctuations. If we denote $E^\pm=\frac{1}{4}\int ({\bf v}\pm {\bf b})^2\, d^3 x$, which are the
  energies carried by Alfv\'en fluctuations propagating in opposite directions (the details are given below), 
  the total energy and cross-helicity of the system are given by $E=E^++E^-$
  and $H_c=E^+-E^-$, respectively. Therefore, when the latter is nonzero,
  the energies of waves propagating along and against the guide
  field are not equal, and in that sense the turbulence is called
  \emph{imbalanced}. 

  Independent numerical simulations by different groups have
  demonstrated that strong MHD turbulence is always locally imbalanced, in the
  sense that in the steady state, the turbulence develops correlated
  regions of positive and negative cross-helicity, irrespective of the
  overall cross-helicity of the system.  Imbalance turbulence is also
  present in the solar wind, where velocity and magnetic fluctuations
  show high correlations of a preferred sign, that is, the normalized
  \emph{cross-helicity} $\sigma_c = {\left\langle\vec v\cdot\vec
  b\right\rangle}/{E}={H_c}/{E}$ is close to unity. The preferred
  positive sign of $\sigma_c$ indicates that there is more energy in
  Alfv\'en waves propagating outwards from the Sun than propagating
  inwards~\cite{belcher71}.
  
  Recently, several phenomenological models have addressed steady-state
  strong imbalanced MHD
  turbulence~\cite{lithwick07,beresnyak08,chandran08,perez09,podesta10}, 
  some with support from numerical simulations and
  observations. However, these models have lead to conflicting
  predictions. For instance, the theories by Lithwick et al \cite{lithwick07} and Beresnyak and Lazarian~\cite{beresnyak08} conclude 
  that in imbalanced regions the Els\"asser spectra have the same
  scalings $E^+(k_\perp)\propto E^-(k_\perp)\propto k_\perp^{-5/3}$. The
  theory by Chandran~\cite{chandran08} proposes that the spectra of $E^+(k_\perp)$ and
  $E^-(k_\perp)$ are different depending of the degree of imbalance.  
  Finally, the analysis by Perez
  \& Boldyrev~\cite{perez09} and Podesta \& Bhattacharjee~\cite{podesta10} finds that the spectra of
  $E^+(k_\perp)$ and $E^-(k_\perp)$ have different amplitudes but the
  same scalings $E^+(k_\perp)\propto E^-(k_\perp)\propto
  k_\perp^{-3/2}$.

  In this work we present results from high resolution direct numerical
  simulations of strong and steadily driven MHD turbulence, aimed to
  elucidate the role of cross-helicity and resolve the controversies
  among different theories.  Hereafter, strong turbulence is defined
  in the sense of Golreich and Sridhar~\cite{GS95}.   
  This paper is organized as follows. Section
  \ref{modeleqns} briefly describes the MHD equations and the
  relevance of using the Reduced MHD model for strong turbulence
  simulations. Section \ref{numstrategy} describes in detail the
  proposed numerical strategy and discusses important numerical
  aspects that need to be considered when simulating strong imbalanced
  MHD turbulence. In section \ref{numresults}, the most relevant
  results from an extensive number of numerical simulations are
  presented. In section \ref{pheno} a phenomenological model for the
  simulation results is proposed and section \ref{conclusion} presents
  the conclusions.

  \section{Model equations}\label{modeleqns}
In the presence of a guide field ${\bf B_0}$, the incompressible MHD
  equations describing the evolution of magnetic and velocity
  fluctuations, ${\bf b}(\vec x,t)$ and ${\bf v}(\vec x,t)$, can be
  written in terms of the so-called Els\"asser variables~\cite{elsasser56}, ${\bf
  z}^{\pm}={\bf v}\pm {\bf b}$:
  \begin{equation}
    \(\frac{\partial}{\partial t}\mp{\bf v}_A\cdot\nabla\){\bf
      z}^\pm+\left({\bf z}^\mp\cdot\nabla\right){\bf z}^\pm = -\nabla
      P+\nu\nabla^2\vec z^\pm+{\bf f}^{\pm},
    \label{mhd1}
  \end{equation}
  where ${\bf v}_{A}={\bf B}_{0}/\sqrt{4\pi \rho}$ is the Alfv\'en
  velocity, $\rho$ is the fluid density, $P$ is the total pressure that
  is determined from the incompressibility condition, $\nabla \cdot {\bf
  z}^{\pm}=0$, ${\bf f^\pm}$ are large-scale forcing, and $\nu$ is the
  viscosity which acts as an energy sink at small scales.  In contrast with a uniform flow in hydrodynamic, which can be removed by a Galilean transform, the guide magnetic field, whether imposed
  externally or created by large-scale fluctuations, is essential in MHD turbulence. It cannot be
  removed by Galilean transform and it
  mediates nonlinear interactions at all smaller scales. The linear
  terms, $({\bf v}_A\cdot\nabla) {\bf z}^\pm$, describe advection of
  Alfv\'en wave packets along the guide field, while the nonlinear
  interaction terms, $\left({\bf z}^\mp\cdot\nabla\right){\bf z}^\pm$,
  are responsible for energy redistribution over scales. It can be
  observed from equations \eqref{mhd1} that nonlinear interactions can 
  only occur between $\vec z^+$ and $\vec z^-$, and such interactions
  take place when the fields overlap or ``collide'' with each other. It is convenient to describe the resulting energy spectra in terms of the so-called Els\"asser energies, defined as $E^{\pm}=\int |{\bf z}^{\pm}|^2 d^3 x/4$. 

  In the limit of small amplitude of fluctuations, the 
  incompressible MHD system~\eqref{mhd1} describes non-interacting
  linear Alfv\'en waves with dispersion relation $\omega^\pm(\vec k)=\pm
  k_\| v_A$. The incompressibility condition requires these waves to be
  transverse, and they are typically decomposed into two polarizations,
  the shear-Alfv\'en wave ($\vec z^\pm_{S}$) and the pseudo-Alfv\'en
  waves ($\vec z^\pm_{P}$) given in Fourier space
  \begin{eqnarray}
  \vec e_S\equiv \frac{\vec e_\| \times\vec k}{k_\perp}, \quad\label{shear} 
  \vec e_P\equiv \frac{\vec k\times\vec e_S}k, 
  \end{eqnarray}
  where ${\bf e}_\|$ is the unit vector in the direction of the guide field. 
  Strong MHD turbulence is dominated by fluctuations with $k_\perp \gg
  k_\|$.  \citet{GS95} argued that since for large $k_{\perp}$ the
  polarization of the pseudo-Alfv\'en fluctuations is almost parallel to
  the guide field, such fluctuations are coupled only to field-parallel
  gradients, which are small since $k_\| \ll k_{\perp}$. Therefore, the
  pseudo-Alfv\'en modes do not play a dynamically essential role in the
  turbulent cascade. We can remove the pseudo-Alfv\'en modes by setting
  $\vec z^\pm_\|=0$ in equations~\eqref{mhd1} to obtain
  \begin{equation}   
    \displaystyle \partial_t {\tilde {\bf z}}^{\pm} \mp ({\bf V}_A\cdot
    \nabla){\tilde {\bf z}}^{\pm}+({\tilde {\bf z}}^{\mp}\cdot
    \nabla){\tilde {\bf z}}^{\pm}=-\nabla_{\perp} P +\frac
    1R\nabla^2{\tilde {\bf z}}^{\pm},
    \label{rmhd} 
  \end{equation}
  where $R$ denotes the Reynolds numbers (discussed below).  In this
  system, the fluctuating fields have only two vector components,
  ${\tilde {\bf z}}^{\pm}=\{{\tilde z}^{\pm}_1, {\tilde z}^{\pm}_2, 0
  \}$ (where we chose the $z$ axis along the guide field $B_0$) 
  but depend on all three spatial coordinates.  It can be
  demonstrated that system (\ref{rmhd}) is equivalent to the Reduced MHD
  model, originally developed for tokamak plasmas by \cite{kadomtsev74}
  and \cite{strauss76}.

  \section{Numerical approach}
  \label{numstrategy}
  The reduced MHD model ~(\ref{rmhd}) describing the nonlinear
  interactions of Shear-Alfv\'en modes is a valuable tool in theoretical
  and numerical studies of incompressible MHD turbulence. Based on this
  model, we now discuss the conditions that should be satisfied in order
  to correctly simulate strong MHD turbulence, both balanced and
  imbalanced.

  \subsection{Computational domain}

  In the presence of a strong guide field, MHD turbulence is inherently
  anisotropic. It is important to point out that such anisotropy will
  develop deep in the inertial range even if it is not present at the
  outer scale where the turbulence is driven. This can be understood
  from the Golreich and Sridhar~\cite{GS95} picture of strong
  turbulence. As the turbulence cascades to smaller scales, eddies
  become more and more shrunk in the field-perpendicular direction until
  fluctuations satisfy the critical balance condition $k_{\|}B_0\sim
  k_\perp b_{\lambda}$, where $k_\|\sim 1/l$ and $k_\perp\sim 1/\lambda$
  are the field-parallel and field-perpendicular wave vectors associated
  with an anisotropic eddy of parallel size $l$ and perpendicular size
  $\lambda$, respectively, and $b_\lambda$ denoted rms magnetic fluctuations at scale $\lambda$. It is assumed that $b_\lambda\sim v_\lambda$. 

  In a traditional cubic simulation box, when an anisotropic wave packet
  fits the field-parallel direction ($z-$dimension of the box) its
  field-perpendicular dimensions are much smaller than the $x$ and $y$
  dimensions of the box.  This is obviously not an optimal situation,
  since the field-perpendicular resolution required to resolve such a
  wave packet should be much higher than the field-parallel
  resolution. This decreases the effective field-perpendicular
  resolution of the simulations. For example, in the case a cubic
  $512^3$ box with the guide field $B_0=5 b_{rms}$, one will have an equivalent
  field-perpendicular resolution of only $512/5 \sim 100$.

  For our simulations, we define the nonlinear interaction strength
  parameter
  \begin{equation}
    \chi=(k_\perp  b_{\lambda})/(k_\|B_0),
  \end{equation}
  so that the critical balance condition then implies~$\chi\sim 1$.

  Simulations of steadily-driven (dissipative) incompressible
  turbulence are generally based on the Fourier pseudo-spectral
  method. As with any spectral method, the solution to the
  differential equation is approximated by a truncated Fourier
  expansion, and the partial differential equation is converted to a
  set of ordinary differential equations in time for the $N$ Fourier
  coefficients. Nonlinear terms in this representation become
  convolutions whose direct computation requires $O(N^2)$ operations.
  In pseudo-spectral methods, the convolutions are computed in real
  (or configuration) space by 
  means of a Fast Fourier Transform (FFT) that only requires $O(NlogN)$
  operations.  For the numerical
  simulations presented in this work, a third order semi-implicit
  Runge-Kutta/Crank-Nicholson method was used for the time integration. 

  In order to allow for the inertial 
  interval to develop, turbulence is driven at the lowest resolvable
  wave numbers, and the energy dissipates at large wave numbers
  determined by the Reynolds numbers. For simulations on a cubic
  periodic box of size $L$, the smallest wave-numbers along the
  field-parallel and field-perpendicular directions coincide, i.e.,
  $k_\perp = k_\|=2\pi/L$. Therefore, driving at the low $k_\|,k_\perp$
  results in an isotropic forcing and the nonlinear strength parameter
  at the forcing scale becomes $ \chi_0 = {k_\perp b_{\lambda}}/{k_\|B_0}\sim {b_{\lambda}}/{B_0}\ll 1 $, which means that
  at least at the large scales, nonlinear interactions are weak. This
  would not be harmful if we had the resources to achieve arbitrarily
  high resolution, as the turbulence would proceed weakly until
  $\chi\sim 1$, and then it would become strong. However, simulations
  generally produce a rather limited inertial range, so that the
  parameter $\chi$ can hardly reach unity in such a setup.

  As pointed out by \cite{maron01}, and applied in recent
  simulations~\cite{mason06,mason08,perez08}, an effective way to avoid this
  is to use an anisotropic domain such that at the forcing scale the
  parameter $\chi$ is already of order unity, that is, the excited
  large-scale modes are already anisotropic and satisfy the critical
  balance condition.  To achieve this we choose an elongated box
  $L_\perp^2\times L_\|$, so that the lowest field-perpendicular and
  field-parallel wave-numbers are $k_\perp = 2\pi/L_\perp$ and
  $k_\|=2\pi/L_\|$, respectively. In this case, forcing at the lowest
  $k_\perp,k_\|$ leads to $\chi_0 = (L_\| b_{\lambda})/(L_\perp
  B_0), $ which is of order unity provided that
  \begin{equation}
    {L_\perp}/{L_\|}\sim { b_{\lambda}}/{B_0}.
    \label{crit_bal}
  \end{equation}
  In this way, the turbulence is excited in a strong regime and the
  cascade proceeds down to smaller scales preserving the critical
  balance condition.

  \subsection{Numerical Resolution}
  At first sight, it appears that elongating the box along the $z$
  direction to match the elongation of the eddies should not change the
  number of grid points required in this direction compared to the
  number of points in the $x$ direction. Fortunately, the number of points
  in the $z$ direction can be reduced. This follows from the fact that the
  turbulent spectrum declines quite slowly, as a power-law, in the
  $k_\perp$ direction, while it drops sharply in the $k_\|$ direction
  for $k_\|> k_{\perp}^{\alpha}$, where $\alpha$ is a some positive
  power not exceeding~1 \citep{cho00,maron01,oughton04,perez08,perez09}
  . This qualitatively different spectral behavior in $k_\|$ and
  $k_\perp$ directions allows one to reduce the numerical resolution by
  a factor of 2 to 4 in the parallel direction, see
  Table~{\ref{sims_table}}. We checked that the restoration of the full
  resolution in the z direction does not change the results, while
  significantly increases the computing costs.

  \subsection{Periodic Boundary Conditions} The spectral method assumes
  periodic boundary conditions in all spatial directions. The periodic
  boundary conditions in the direction of the guide field (the $z-$direction) 
  may raise the question of whether the magnetic field lines are
  periodic in such numerical simulations. If they were periodic then  
  the Alfv\'en modes counter-propagating along a given magnetic field
  line would repeatedly interact only among themselves, and, therefore, they might not  
  become sufficiently decorrelated between the consequent interactions. To answer  
  this question we note that the numerical setup ensures periodicity of the
  fluctuations ${\bf b}$, but not the magnetic field lines. Magnetic field lines 
  are integral lines of a magnetic field, therefore, they are  
  generally not periodic in the $z-$direction, 
  since ${\bf b}({\bf k}_\|=0)$ is generally nonzero. Consequently, an eddy traveling along a magnetic
  field line interacts with  
  many independent counter-propagating 
  eddies. Reducing the parallel box size
  $L_\|$ may however lead to artificial overlap of 
  an elongated eddy with itself. This may explain why reducing the parallel box size
  $L_\|$ below (\ref{crit_bal}) somewhat spoils the spectrum at low wave
  numbers, as seen in forced simulations of M\"uller \& Grappin~\citep{muller05}. 
  Increasing $L_\|$ beyond (\ref{crit_bal}) does not change the
  results but increases the computational cost [Mason \& Cattaneo,
  private communication 2006].
  
\subsection{The strength of the guide field}
  In the inertial interval of turbulence, the guide field should be
  strong compared to the turbulent fluctuations, $b_\lambda \ll B_0$. It is
  important to establish how small the  fluctuations should be in order
  to exhibit the universal turbulent spectrum. The question is
  especially relevant for numerical simulations, as the large guide
  field implies small Alfv\'enic time and, therefore, large computing
  cost. This problem was numerically addressed in \cite{mason06}, where
  it was found that the transition to the universal regime occurs
  approximately at $b_\lambda/B_0\sim 1/3$. For $b_\lambda/B_0 >1/3$ the energy
  spectrum is closer to $-5/3$, possibly indicating that the magnetic
  field does not qualitatively change the Kolmogorov dynamics for the scales attainable in the numerical simulations. 
  For $b_\lambda/B_0 <1/3$, the guide field significantly
  affects the dynamics, and the spectrum changes to $-3/2$. The ``sweet
  spot" often used in numerical simulations is $b_{rms}/B_0\sim 1/5$; it
  has been checked that the smaller ratio, $b_\lambda/B_0\sim 1/10$, does not
  lead to noticeably different
  results (see for example ~\cite{muller05,mason06,mason08}).   
  
  It is also important to note that in numerical simulations, where
  inertial intervals have quite limited extent, the condition
  $b_{rms}/B_0<1/3$ should be satisfied at the outer scale. Indeed, since
  $b_\lambda$ decreases with the scale quite slowly, say $b_\lambda
  \propto \lambda^{1/3}$, if the condition $b_{rms}/B_0<1/3$ is not
  satisfied at the outer scale, it can hardly be satisfied in the inertial
  interval.

  \subsection{Reynolds numbers}\label{re} 
  Probably the most significant limitation on numerical simulations 
  is imposed by the consideration of imbalanced turbulence. Indeed, in
  the imbalanced case, $\gamma =z^+/z^-> 1$, the formally constructed Reynolds numbers $Re_\lambda^{\pm}=\lambda  z_\lambda^{\mp}/\nu$ 
  corresponding to $\vec z^+$ and $\vec z^-$ fields at some scale $\lambda$ would be essentially
  different (here $z^\pm\equiv |\vec z^\pm|$). One can argue that   
  the effective Re number is then the smaller of the two, see below. Therefore, the resolution 
  requirements increase with the 
  amount of imbalance in order to produce large inertial ranges. Assume
  that the number of grid points in a field-perpendicular direction
  scales with the Reynolds number as $N\sim Re^{\beta}$, where
  $\beta=2/3$ or $3/4$ depending on the spectral slope ($3/2$ or
  $5/3$). Then increasing the imbalance $\gamma$ by, say, a factor of 3
  will require increasing the resolution by approximately a factor of
  2. Noting that resolution of at least 1024 is required to simulate the
  imbalance $\gamma\sim 2$ (see below), we conclude that significantly
  stronger imbalance is not achievable with present day computing power.

  \subsection{Random Forcing} 
  Another question that arises in the strong turbulence regime
  concerns the type of forcing.  In real systems, large-scale turbulence can be driven by
  instabilities exciting both velocity and magnetic harmonics, by
  external antennae exciting magnetic fields, etc.; this should not matter for the spectrum of 
  small-scale fluctuations. Similarly, in numerical simulations one can force at large 
  scales either velocity or magnetic field, or both of them, this does not change the result~\citep{mason08}. The goal is therefore not to mimic any real-life driving but rather to optimize the transition to the inertial interval. 

  In this section we discuss the important aspects to be considered when
  choosing a particular forcing. We assume a random force ${\tilde {\bf
  f}}$ that has no component along $z$, it is solenoidal in the $x-y$
  plane and its Fourier coefficients outside the range
  \begin{equation}
    1 \leq k_{\perp} \leq 2, \quad(2\pi/L_\|) \leq k_\| \leq
    (2\pi/L_\|)n_z\label{modes}
  \end{equation}
  are zero, where $n_z$ determines the width of the force spectrum in
  $k_\|$, and $L_\perp = 2\pi$. The Fourier coefficients inside that
  range are Gaussian random numbers with amplitude chosen so that the
  resulting rms velocity fluctuations are of order unity.  The
  individual random values are refreshed independently at time
  intervals~$\tau$.  The parameter~$n_z$ controls the degree to which
  the critical balance condition is satisfied at the forcing scale. Note
  that we do not drive the $k_\|=0$~mode but allow it to be generated by
  nonlinear interactions.

  \begin{figure}[!t]
    \includegraphics[width=0.5\textwidth]{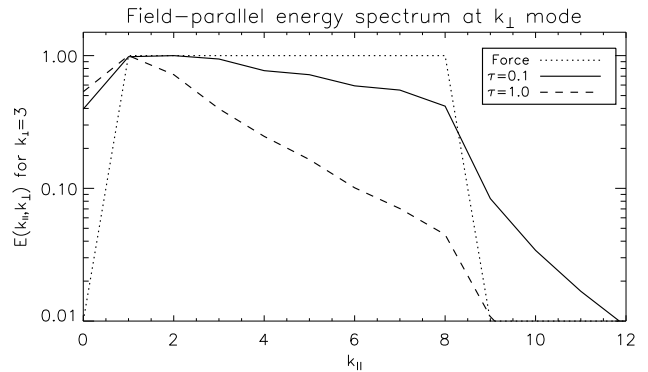}
    \caption{Field-parallel energy spectrum at the dominant mode
      $k_\perp =3$ for different forcing correlation times. The dotted
      line represents the normalized spectrum of the forcing. We observe
      that as the correlation time increases, large $k_\|$ modes get
      suppressed despite the fact that force amplitude is the same for
      all modes.}
    \label{response}
  \end{figure}

  In contrast with incompressible hydrodynamic system, an incompressible
  MHD system can support Alfv\'en waves. When the system is driven by a
  time dependent forcing, the most effectively driven modes are those
  resonating with the frequencies present in the forcing. Therefore, the
  spatial spectra of the large-scale velocity and magnetic fluctuations
  are not generally the same as the spatial spectrum of the force.
  Rather, they essentially depend on the {\em both} spatial {\em and}
  temporal spectra of the random forcing.  Ultimately, it is the
  spectrum of the large-scale velocity and magnetic fluctuations, not the driving force, that
  should be controlled in numerical simulations. To illustrate this we
  performed a series of simulations in which we drive the modes given by
  equation \eqref{modes} with $n_z=8$. We would expect this to excite
  all modes from $k_{0\|}=(2\pi/L_z)$ to $8k_{0\|}$. Figure
  \ref{response} shows the field-parallel energy spectrum of
  fluctuations at $k_\perp=3$, for differing values of the force
  correlation time, showing that for short time correlations, the fluid 
  response broadens.
 
  This adds additional complexity as the way in which the turbulence is
  forced may affect the outcome of the simulation's results. In MHD
  turbulence there are regimes where the spectrum is expected to obey
  universal power laws, as well as transition regions where the
  turbulence changes character, for instance from weak to strong
  turbulence. Therefore, when simulating MHD turbulence, meaningless
  results could easily be obtained if the forcing is not chosen
  carefully.  Not optimized forcing makes the transition from the forcing scale to the
  inertial range unnecessarily longer, as the turbulent fluctuations have
  to develop the anisotropy implied by critical balance.  In
  the following we perform simulations with a short-time-correlated
  random forcing that drives the turbulence close to the critical
  balance condition at the large scales. 
  A short-time correlated Gaussian random force has another
  important advantage, that it allows one to control the rates at which the
  energies of $z^+$ and $z^-$ modes are injected.

  \subsection{Inertial range and bottleneck} 
  In spite of the significant growth in massively parallel
  supercomputers that has occurred over the last years, simulations of
  MHD turbulence still result in a rather limited inertial range. This
  generally raises the question of as to whether the simulations fully
  capture all scales from forcing to dissipation and whether the scaling
  exponents inferred from the energy spectrum are accurate enough to
  confirm or rule out the existing models of turbulence. In order to
  extend the inertial range, some groups have performed simulations with
  hyper-viscosity of different orders. However, using artificial
  viscosity might enhance possible bottlenecks (or bumps) that arise at
  high $k$ due to an abrupt viscous suppression of the turbulent cascade
  in the dissipation range, e.g., \cite{borue95,cho00}. This bottleneck can significantly affect
  measured scaling laws in simulations with inertial ranges of limited
  extent. We avoid the use of hyper-viscosity and perform direct
  numerical simulations of the RMHD equations \eqref{rmhd}.

  In our results we identify the inertial range by performing a set of
  simulations from low to high Reynolds numbers (with increasing
  resolution). It is verified that the inertial range becomes larger as
  the Reynolds number increases. At larger $k$ the inertial interval is
  followed by the dissipation range where the power-law scaling does not
  apply.

  Spectral power laws in the simulations are detected by compensating
  the energy spectra with corresponding power laws. This method is an efficient
  way to compare simulations with existing theories and to observe any
  deviation from the expected scalings. As another advantage of 
  compensating the energy spectrum, one notices that since the compensating
  factor increases with $k_\perp$, it enhances any possible
  bottleneck region occurring at large $k_\perp$. In the numerical
  results that we present in this paper,  no 
  evidence of a significant
  bottleneck effect is found, which is consistent with simulations by other groups, e.g.,  \cite{maron01,muller05,mason06}.  Note that for the resolution used in our
  paper, the corresponding HD simulations would already produce well developed and clearly observed bottleneck region~\cite{gotoh02}.

  \subsection{Locality}
  The question of universality of MHD turbulence is closely related to
  the question of locality of nonlinear interactions. Locality loosely
  implies that only fluctuations of comparable length scales interact strongly
  with each other. More precisely, locality means that the properties of
  the inertial interval asymptotically become independent of the details
  of the driving and the dissipation as the Reynolds number
  increases. Locality is an essential property of hydrodynamic
  turbulence. In relation to MHD turbulence, it has been discussed in several works, e.g.,  \cite{alexakis05,alexakis07,carati06,yousef07}. Recently, Aluie and Eyink~\cite{aluie10} showed both analytically and numerically that, similarly to hydrodynamic turbulence, scale-locality holds in MHD turbulence. This is consistent with the fact
  that numerical simulations of strong MHD turbulence performed with different forcing and dissipation mechanisms
  produce the same energy spectrum, e.g., \cite{maron01,muller05,mason06,mason08,perez08}.   
   
  \subsection{Integration time}
  Finally, we point out that numerical simulations of imbalanced MHD
  turbulence require longer integration time in order to accumulate good
  statistics.  This may be related to the fact that the nonlinear
  interaction is reduced in imbalanced turbulence, see
  subsection~\ref{re}. The numerical simulations presented in the next
  section indicate that for a modest imbalance of about $\gamma \sim
  2-3$, the relaxation time of the turbulence spectrum is about 20 formally estimated 
  large-scale dynamical times (see below), while about 100 dynamical times are
  required to  accumulate good statistics. Numerical simulations with significantly 
  shorter integration time do not produce reliable results.

  \section{Numerical Results}\label{numresults}

\begin{figure}[!htbp] 
    \includegraphics[width=0.5\textwidth]{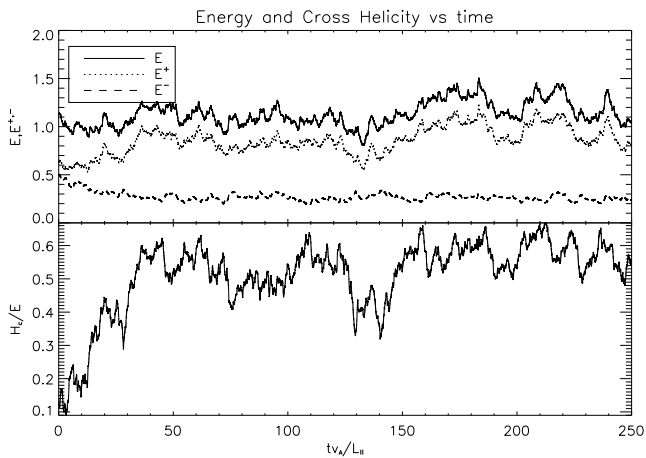}
    \caption{Evolution of energy for imbalanced run. The outer-scale dynamic time is about $(L_\perp/3)/u_{rms} \sim 2$.}
    \label{energy}
  \end{figure}

  Equations \eqref{rmhd} are evolved until a stationary
  state is reached, as determined by the time evolution of the total
  energy of the fluctuations, (see figure \ref{energy}). A
  typical run produces over 200 snapshots; the large-scale dynamic time
  associated to the dominant large scale mode $k_\perp\sim 3$ is
  $(L_\perp/3)/u_{rms}\sim 2$.     

\begin{figure}[!th]
    \begin{center}
      \includegraphics[width=0.5\textwidth]{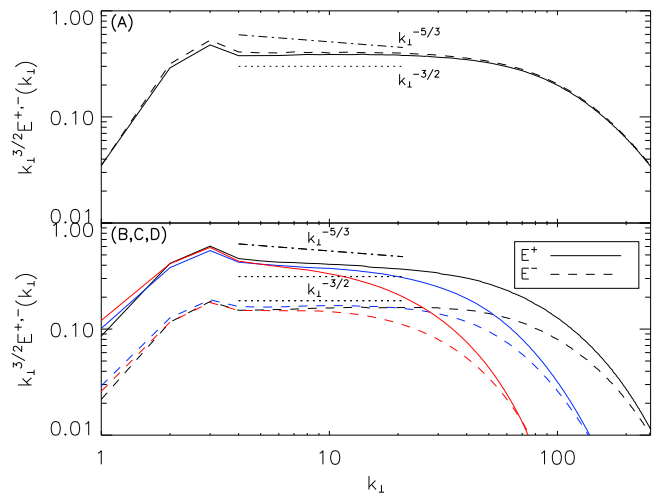}
    \end{center}
    \caption{Spectra of the Els\"asser fields in numerical simulations of strong MHD turbulence. Top two
      frames: balanced turbulence (run A); bottom frame: imbalanced
      turbulence (runs B(red), C(blue) and D(black)).}\label{spectra}
  \end{figure}

  Since the background magnetic field must be strong, we choose $B_0=5$
  in the $v_{rms}$ units, (see the discussion in~\citep[][]{mason06}). Time
  is normalized to the Alfv\'en transit time $\tau_0=L_\|/v_A$,
  where $L_\|$ is the
  field-parallel box size. This time is equivalent to the
  perpendicular transit time $L_\perp/v_{rms}$ when $v_{rms}=1$. The
  Reynolds number is defined as 
  $Re=v_{rms}(L_\perp/2\pi)/\nu$ and we have chosen the same value for
  the magnetic Reynolds number, $Rm={b}_{rms}(L_\perp/2\pi)/\eta$,
  denoting both by~$R$ in~(\ref{rmhd}). In each run, the average is
  performed over about 100 large-scale-eddy turnover times. The results
  are presented in Fig.~(\ref{spectra}).

    \begin{table}[!tb]
    \begin{tabular}{c@{\hspace{0.5cm}}c@{\hspace{0.5cm}}c@{\hspace{0.5cm}}c@{\hspace{0.5cm}}c@{\hspace{0.5cm}}c@{\hspace{0.5cm}}c@{\hspace{0.5cm}}} \hline\hline 
      Run & Resolution & $Re$ & $\nu=\eta $ & $L_\|/L_\perp$ &
      $\sigma_c$ \\ \hline A & $ 512^2\times 256$ & 2400 &$4.2\times
      10^{-4}$ & 5 & 0 \\ B & $ 256^3$ & 900 &$1.1\times 10^{-3}$ & 10
      & 0.6 \\ C & $ 512^2\times 256$ & 2200 &$4.6\times 10^{-4}$ &10 &
      0.6 \\ D & $1024^2\times 256$ & 5600 &$1.8\times 10^{-4}$ &10 &
      0.6 \\ \hline\hline
    \end{tabular}
    \caption{Summary of simulations of strong balanced turbulence (A)
      and strong imbalanced turbulence (B, C, D).}\label{sims_table}
  \end{table}
  
  Table \ref{sims_table} summarizes five representative simulations that
  incorporate all the aspects discussed in section \ref{numstrategy}. 
  Run  A correspond to
  balanced turbulence, that is, $\sigma_c=0$. In runs labeled B, C and D, cross
  helicity is injected at the forcing scale in such a way that $\sigma_c$ 
  reaches a steady state of $\sim 0.6$. We
  use a short time-correlated forcing, which is on average $1/20^{th}$
  of the Alfv\'en time of
  the excited modes, so that the energy injection rates for both $\vec
  z^+$ and $\vec z^-$ only depend on the variance of the imposed forcing,
  which is controlled in our simulations. In the imbalanced
  case, field-parallel box size is optimized to reach the critical balance at the
  large scales. Except for the Reynolds numbers, simulations B, C, and D
  have the exact same parameters including the energy injection rates,
  $\epsilon^+$ and~$\epsilon^-$.

  The field-perpendicular energy spectrum is obtained by 
  averaging the angle-integrated Fourier spectrum,
  \begin{equation}
    E(k_\perp)=0.5\langle|{\bf v}({\bf 
      k_\perp})|^2\rangle k_\perp+0.5\langle|{\bf b}({\bf
      k_\perp})|^2\rangle k_\perp,
  \end{equation}
  over field-perpendicular planes in all snapshots.
  Figure~\ref{spectra} shows the field-perpendicular energy 
  spectra  for each run.  

  Our numerical setup offers significant advantages over full MHD
  simulations, and can be already seen in simulations 
  of balanced turbulence, top frame in Fig.~(\ref{spectra}), run
  A. The energy spectra approach $E^\pm(k_\perp)\propto
  k_\perp^{-3/2}$, 
  in good agreement with earlier numerical findings
  \citep[e.g.,][]{maron01,haugen03,muller05,mason06}, but requiring
  considerably less computational cost and producing 
  slightly larger 
  inertial intervals. Our most significant results are obtained for
  the imbalanced case.  The bottom frame of Fig.~\ref{spectra} shows the
  spectra for three different Reynolds numbers (runs B, C, D). It is
  observed that the spectra $E^\pm$ are pinned at the
  dissipation scales, which supports the phenomenological predictions
  by \cite{grappin83,galtier00,lithwick03,chandran08}. We also find that the
  large-scale parts of the both spectra are practically insensitive to the
  Reynolds numbers. These two important properties imply that as the Re
  numbers are further increased, the $E^\pm$ spectra must become
  progressively more parallel in the inertial interval. This is indeed
  seen in our numerical simulations (B, C, D). Our numerical
  simulations indicate that both spectra approach the universal scaling of
  strong MHD turbulence $E^\pm(k_\perp)\propto k_\perp^{-3/2}$, while they
  have essentially different amplitudes and correspond to essentially
  different energy fluxes.

  \section{Phenomenological modeling}\label{pheno}
\begin{figure}
\vspace*{-0.5cm}
    \begin{center}
      \includegraphics[width=3.6in]{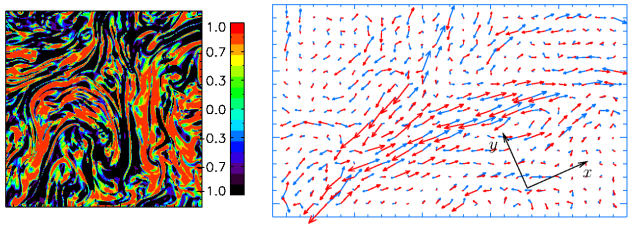}
    \end{center}
    \vspace*{-0.5cm}
    \caption{Left: Cosine of the alignment angle between ${\bf v}_{\lambda}$
      and ${\bf b}_{\lambda}$ fluctuations in the guide-field
      perpendicular plane at scale
      $\lambda=L_\perp/12$ in Run A. Right: A correlated region of (counter-)aligned magnetic and
      velocity fluctuations (red and blue vectors) at scale $\lambda=L_\perp/12$, in a
      plane perpendicular to the strong guide field, in run A. The fluctuations are aligned 
      predominantly in the $x$~direction while their directions and amplitudes change 
      predominantly in the~$y$ direction.}\label{domains_corr}
  \end{figure}

  In this section we propose an explanation for the observed spectra. In the case 
  of balanced MHD turbulence, our explanation of the $k_\perp^{-3/2}$ energy spectrum relies on the phenomenon of \emph{scale-dependent} dynamic alignment, see~\cite{boldyrev06,mason08}. (General
  details on dynamic alignment in MHD turbulence can be found in,
  e.g.,~\cite{matthaeus84,biskamp}).  Consider the eddy shown
  to the right of Fig.~\ref{domains_corr}, obtained from simulations. In this eddy
  fluctuations are aligned within the small angle $\theta_{\lambda}$
  along $x$, while their directions and magnitudes change in an almost
  perpendicular direction, along $y$.  In the case of strong balanced
  turbulence, the nonlinear interaction in such an eddy is then
  reduced by a factor $\theta_{\lambda}$ for both $z^+$ and $z^-$
  fields, and the corresponding nonlinear interaction time is
  estimated as $\tau_{\lambda} \sim 1/({\bf z}^{\pm}_{\lambda} \cdot
  {\bf k}_{\perp}) \sim 1/(z^{\pm}_{\lambda}k_{\perp}
  \theta_{\lambda})$. The scaling of the fluctuating fields is then
  found from the requirement of constant energy fluxes:
  $(z_{\lambda}^{\pm})^2/\tau_{\lambda}={\rm const}$.  One can argue
  \citep{boldyrev06,mason06,mason08} that the alignment angle
  decreases with scale as $\theta_{\lambda}\propto \lambda^{1/4}$, in
  which case the field-perpendicular energy spectrum is
  $E(k_{\perp})\propto k_{\perp}^{-3/2}$.
  \begin{figure}
    \centering \includegraphics[width=1.8cm,
      angle=-90]{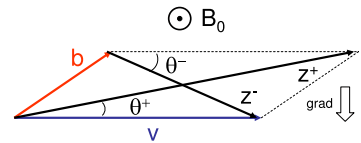}
    \caption{Sketch of dynamic alignment of magnetic and velocity
      fluctuations in a turbulent eddy.}
    \label{fig:angles}
  \end{figure}

  It turns out that the imbalanced simulations of
  previous section can also be explained within the framework of the {\em scale-dependent} dynamic alignment~\cite{perez09,perez10}. Let us assume that the
  alignment is present in the imbalanced case. Since the fields amplitudes $z^+$ and $z^-$ 
  are now essentially 
  different, the alignment angles are different as well; we denote them
  $\theta_{\lambda}^+$ and $\theta_{\lambda}^-$, see Fig.~\ref{fig:angles}. 
  These angles obey the important geometric constraint:
  $\theta_{\lambda}^+z_{\lambda}^+\sim \theta_{\lambda}^-z_{\lambda}^-$,
  as is clear from Fig.~\ref{fig:angles}.  The depletion of nonlinear
  interaction~\cite{kraichnan88,servidio08} is therefore different for $z^+$ and $z^-$ fields, which makes their nonlinear interaction times, $\tau^{\mp}_{\lambda} \sim
  1/(z^{\pm}_{\lambda}k_{\perp} \theta^{\pm}_{\lambda})$, the
  same. The requirement of constant energy fluxes
  $(z_{\lambda}^{\pm})^2/\tau^{\pm}_{\lambda}\sim \epsilon^{\pm}={\rm
    const}$ then ensures that $z_{\lambda}^+/z_{\lambda}^-\sim
  \sqrt{\epsilon^+/\epsilon^-}$, so both fields should have the same
  scaling, although different amplitudes. 
  We also note that the Reynolds numbers that take into account the depletion of nonlinear interaction are the same for both fields, $Re_\lambda^{\pm}\sim \lambda z_\lambda^{\mp}\theta_\lambda^{\mp}/\nu$, and they are reduced approximately by the factor $\gamma=z^+/z^-$ with respect to the formal Reynolds number $Re_\lambda=\lambda v_\lambda/\nu$. 
  
  We note that the above relations of the kind $z_{\lambda}^+/z_{\lambda}^-\sim
  \sqrt{\epsilon^+/\epsilon^-}$ hold locally for a particular domain, with given (positive or negative) alignment, while MHD turbulence consists of both positively and negatively aligned domains of various strengths. Since $E^+$ and $\epsilon^+$ are concentrated mostly in positively aligned domains, while $E^-$ and $\epsilon^-$ in negatively aligned ones, 
  the quantities $\langle z^\pm_\lambda \rangle$ and $\langle \epsilon^\pm \rangle$ averaged over the global system should not a priori satisfy the same relations. The difference between the local and global quantities in MHD turbulence should be taken into account when one designs numerical tests. 
  
Recently, Beresnyak \& Lazarian \cite{beresnyak08,beresnyak09} attempted to test the effects of the dynamic alignment in imbalanced turbulence by measuring the relations among \emph{global} quantities $\langle z^\pm_\lambda \rangle$ and $\langle \epsilon^\pm \rangle$. They did not observe the \emph{local} scaling relations and concluded that the dynamic alignment proposed in~\cite{perez09,perez10} is absent. According to the explanation given above, such a conclusion is incorrect. The relations between the global quantities should depend on the details of the distribution of positive and negative domains in the turbulent system, see the discussion in~\cite{podesta10}. For our present purposes we simply need the fact that if each domain  has the same {\em scaling} of the fluctuating fields, the fields averaged over the whole turbulent system will have the same scaling as well.

  
  \section{Discussion.}\label{conclusion}
  We have presented a detailed numerical setup based on the Reduced
  MHD model (RMHD), and results from high resolution simulations of
  balanced and imbalanced MHD turbulence in steady state. We have used
  this numerical set up to address currently existing
  controversies regarding the spectra of imbalanced MHD
  turbulence. The simulations are consistent with the theories and
  observations predicting same scaling for both Els\"asser fields~
  \citep{beresnyak08,lithwick07,perez09,podesta10} and less consistent with the
  models predicting different scalings for $E^{\pm}$,
  \citep[e.g.,][]{chandran08}. The measured scaling
  exponent in the simulations is close to the $-3/2$ supported by
  phenomenological models based on dynamic
  alignment~\cite{perez09,podesta10}. The
  analysis of our simulation results may also explain
  somewhat puzzling numerical findings by \citet{beresnyak08,beresnyak09}, who  
  report different spectra for the $E^{\pm}$ Els\"asser fields, and
  the intersection of the spectra rather than pinning at the
  dissipation scale. According to our results, the explanation might
  lie in the fact that in these simulations the imbalance was typically  
  high, up to $\gamma^2=(z^+)^2/(z^-)^2 \sim 1000$, and,
  therefore, the universal regime of imbalanced MHD turbulence was not
  reached, see our analysis in section~\ref{re}.

  
  The phenomenology of scale-dependent dynamic alignment
  can be applied to explain the observed spectra. In this
  phenomenology, the configuration space splits into eddies (domains) with
  highly aligned and anti-aligned magnetic and velocity fluctuations,
  where nonlinear interactions are reduced, as in left panel of 
  Fig.~\ref{domains_corr}.
  Even when the turbulence is balanced overall it still can be
  imbalanced locally, creating domains of positive and negative
  cross-helicity. In each of these regions the picture of imbalanced
  turbulence presented above applies. When averaged over all the
  regions, the spectra of balanced turbulence are reproduced.

  This work was supported by the U.S. 
  DOE Junior Faculty grant DE-FG02-07ER54932, by the DOE grant DE-SC0001794, and by the NSF Center for Magnetic
  Self-Organization in Laboratory and Astrophysical Plasmas
  at the University of Wisconsin-Madison. 
  High Performance Computing resources were provided by the Texas
  Advanced Computing Center (TACC) at the University of Texas at Austin
  under the NSF-Teragrid Project TG-PHY070027T. 


\end{document}